# Charged excitons and biexcitons in laterally coupled InGaAs quantum dots

Jie Peng and Gabriel Bester[*]

*Max-Planck-Institut fur Festkorperforschung, Heisenbergstr. 1, D-70569 Stuttgart, Germany*

(Dated: December 9, 2010)

We present results of atomistic empirical pseudopotential calculations and configuration interaction for excitons, positive and negative trions ($X^{\pm}$), positive and negative quartons ($X^{2\pm}$) and biexcitons. The structure investigated are laterally aligned InGaAs quantum dot molecules embedded in GaAs under a lateral electric field. The rather simple energetic of excitons becomes more complex in the case of charged quasiparticles but remains tractable. The negative trion spectrum shows four anticrossings in the presently available range of fields while the positive trion shows two. The magnitude of the anticrossings reveals many-body effects in the carrier tunneling process that should be experimentally accessible.

## I. INTRODUCTION

Epitaxial semiconductor quantum dots (QDs) can be designed with an increasing degree of control over their sizes, shapes and alloy compositions[1,2]. Individual structures can be arranged laterally and vertically using controlled self-assembly or substrate patterning, to generate complex arrangements of QDs. This newly gained ability to deterministically produce architectures of QDs is the prerequisite for the implementation of scalable quantum networks, where each QD holds a quantum bit (qubit). Furthermore, impressive advances were made in the manipulation of the realized quantum states; enabled by a fundamental understanding of the QD's electronic and optical properties. For single QDs, different types of quantum states were already manipulated. Early work focussed on the coherent manipulation of exciton qubits[3–8], but due to their rather short coherence times, the attention has shifted to the electron spin. The initial step of the single spin initialization was demonstrated[9–11] and is now part of more advanced spin manipulation experiments[12–16].

For structures with more than one QD, the most advanced prototypes are vertically aligned (stacked) quantum dot molecules (QDMs). These structures have appeared more than a decade ago[17–19] but have developed into a fertile platform for quantum manipulations only recently. One important step was the fabrication of structures designed to allowing either electron or hole tunneling[20]. The observation of the coupling through anticrossings under vertical electric field[21–24] along with theoretical atomistic modeling[25–27] allowed for precise estimates of many relevant coupling energies. Indeed, the understanding of the quantum states in vertically coupled QDMs is rather deep and goes beyond coupling energies. For instance, the existence of an antibonding ground state[28,29] could be demonstrated, the influence of a lateral misalignment of the QDs was investigated[30] and indirect excitons could be recently observed[31]. Success in the area of *control* and *manipulation* was very recently reported for vertically stacked QDMs. Namely, the optical spin initialization over the fine-structure of the excited trion state[11] and the ultrafast control of the entanglement between two electrons spins located in different dots[32]. This represents the first two-qubit operation with QDs.

The history of *laterally* aligned QDMs is younger, with high quality structures appearing only in the last few year[1,2,33–36]. Laterally coupled QDMs are certainly good candidates for applications in quantum information science because of the potential to couple several QDs ("scaling") to form the first building block of a useful device. It is also believed that the degree of external control of individual QDs within a QDM or an array of laterally aligned QDs should be larger than in vertical structures. However, the control until now has been limited due to the difficulty to apply lateral electric fields in order to gate (or tune) the device. Indeed, the geometrical constraints have lead to relatively weak[37,38] applied lateral fields until now. Likewise, the charging of the QDM with extra carriers, as achieved regularly in vertical structures through tunneling from a δ-doping layer, has not yet been achieved in lateral structures. These limitations are not believed to be of fundamental nature but certainly represent technical challenges.